\documentclass[aps,prb,reprint,showpacs,showkeys,superscriptaddress,groupedaddress]{revtex4-1}
\usepackage{graphicx}% Include figure files
\usepackage{dcolumn}% Align table columns on decimal point
\usepackage{bm}% bold math
\usepackage{hyperref}% add hypertext capabilities
\begin{document}

%\preprint{}

\title{Evidences of evanescent Bloch waves in Phononic Crystals}

\author{V. Romero-Garc\'ia}
 \email{virogar1@mat.upv.es}
\author{J.V. S\'anchez-P\'erez}
 \affiliation{Centro de Tecnolog\'ias F\'isicas: Ac\'ustica, Materiales y Astrof\'isica, Universidad Polit\'ecnica de Valencia.}%Lines break automatically or can be forced with \\

\author{S. Casti\~neira-Ib\'a\~nez}
 \affiliation{Dpto. F\'isica Aplicada, Universidad Polit\'ecnica de Valencia.}%Lines break automatically or can be forced with \\

\author{L.M. Garcia-Raffi}
 \affiliation{Instituto Universitario de Matem\'atica Pura y Aplicada, Universidad Polit\'ecnica de Valencia.}%Lines break automatically or can be forced with \\

\date{\today}

\begin{abstract}
We show both experimentally and theoretically the evanescent
behaviour of modes in the Band Gap (BG) of finite Phononic Crystal
(PC). Based on experimental and numerical data we obtain the
imaginary part of the wave vector in good agreement with the
complex band structures obtained by the Extended Plane Wave
Expansion (EPWE). The calculated and measured acoustic field of a
localized mode out of the point defect inside the PC presents also
evanescent behaviour. The correct understanding of evanescent
modes is fundamental for designing narrow filters and wave guides
based on Phononic Crystals with defects.

\end{abstract}


\maketitle

During the past few years, there has been a great deal of interest
in studying propagation of waves inside periodic structures. These
systems are composites made of inhomogeneous distribution of some
material periodically embedded in other with different physical
properties. Phononic crystals (PC) \cite{Sigalas93, Kushwaha94PRB}
are one of the examples of these systems. PC are the extension of
the so-called Photonic crystals \cite{Yablonovitch} when elastic
and acoustic waves propagate in periodic structures made of
materials with different elastic properties. When one of these
elastic materials is a fluid medium, then PC are called Sonic
Crystals (SC) \cite{Martinez95, Sanchez98}.

For these artificial materials, both theoretical and experimental
results have shown several interesting physical properties
\cite{Joannopoulus08}. In the homogenization limit
\cite{Torrent06}, %i.e. in the limit of large wavelengths,
it is possible to design %clusters of cylinders to obtain
acoustic metamaterials %with prefixed parameters mainly determined
%by the fraction of the volume occupied by the scatterers, and they
that can be used to build refractive devices \cite{Torrent07}. In
the range of wavelengths similar to the periodicity of the PC
($\lambda\simeq a$), multiple scattering process inside the PC
leads to the
phenomenon of so called Band Gaps (BG),%. Waves with frequencies
%inside the BG do not propagate through the periodic structure. BG
%are for instance
which are required for filtering sound \cite{Sanchez98}, trapping
sound in defects \cite{Wu01, Wu09} and for acoustic wave guiding
\cite{Vasseur08}.

Propagating waves inside a periodic media are a set of solutions
of the wave equations satisfying the translational symmetry
property. However, periodic media with point defects where the
translational symmetry is broken, or finite periodic media, can
support evanescent modes as well. Recently Laude et al.
\cite{Laude09} have analyzed the evanescent Bloch waves and the
complex band structure of PC. Complex band structures show bands
that are simply not revealed by the traditional $\omega(k)$
method. By means of the complex band structures, BG can be defined
as ranges of frequencies where all Bloch waves must be evanescent.

The goal of the paper is to characterize the evanescent behaviour
of waves with frequencies in the BG inside of PC. Analytical,
numerical and experimental data show the evidences for the
exponential-like decay of these modes. Engelen et al.
\cite{Engelen09} shown that modes inside BG in photonic crystals
decay multiexponentially. Supercell approximation in Extended
Plane Wave Expansion (EPWE) \cite{Hsue05, Laude09, Romero10} has
been used in the preseent work to determine the imaginary part of
the wave vector of evanescent modes. Specifically, we have deduced
both analytically and experimentally
the imaginary part of the wave vector %from the experimental data
observing that only the first harmonic contributes substantially
to the decay of the acoustic field inside complete SC. %We have
%also seen both numerically and experimentally the evanescent
%behaviour of localized modes.
In all cases we have obtained a very good agreement between
theoretical and experimental results.

We have performed experiments in an echo-free chamber of
dimensions 8$\times$6$\times$3m$^3$. The finite 2D SC used in this
paper forms a square array with lattice constant $a=22$cm. The
size of the SC is 5$a\times$5$a$ and the radius of the cylinders
is $r=10$cm. A prepolarized free-field 1/2" microphone Type $4189$
B\&K has been used throughout the experiments. The diameter of the
microphone is $1.32$cm, which is approximately $0.06a$. Our system
3DReAMS (3D Robotized e-Acoustic Measurement System) is capable of
sweeping the microphone through a 3D grid of measuring points
located at any trajectory inside the echo-free chamber. Motion of
the robot is controlled by NI-PCI 7334.

\begin{figure}
\includegraphics[width=85mm,height=50mm,angle=0]{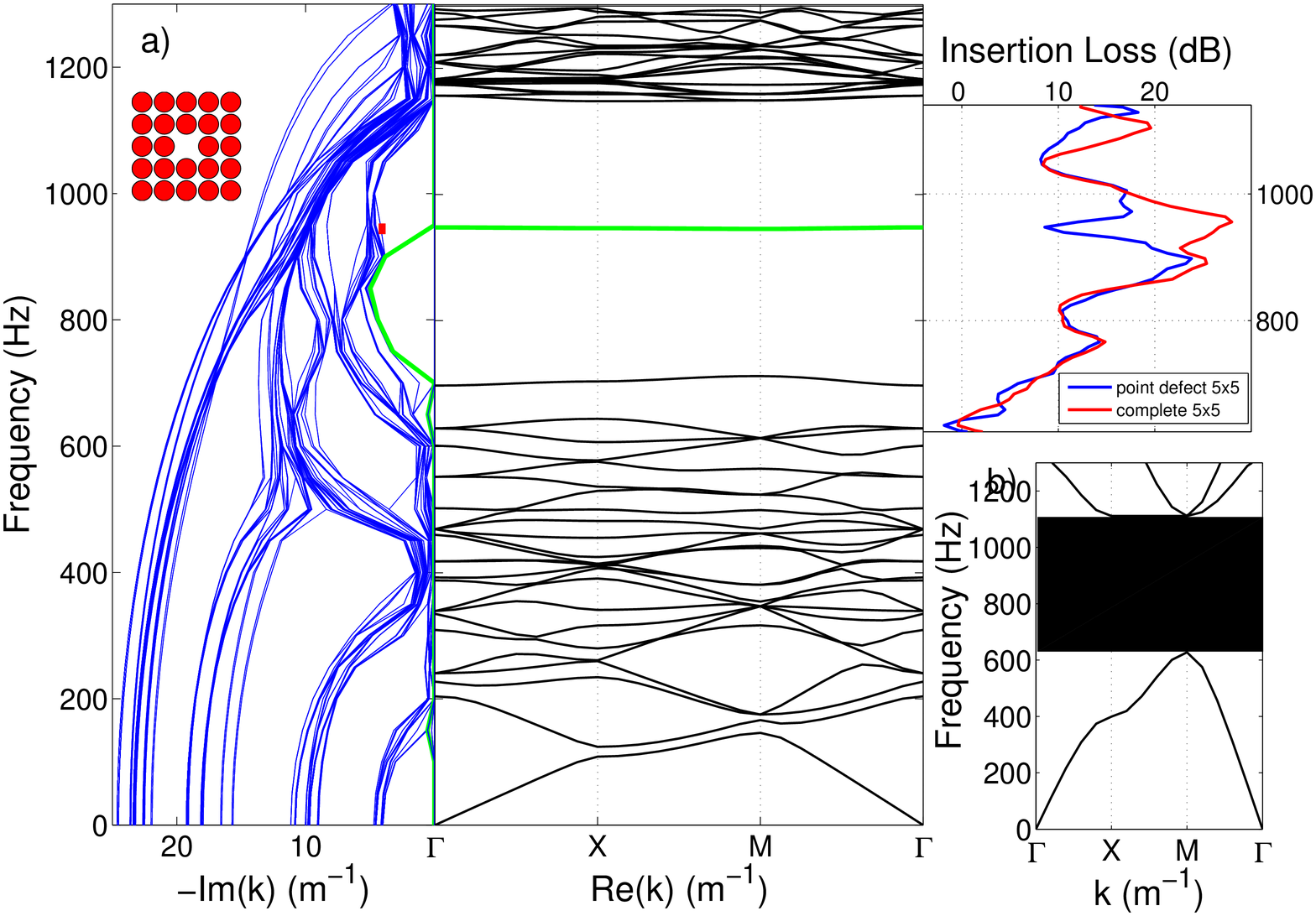}%
\caption{\label{fig:Figure_1}(Color online) Band Structures of a
complete SC and of a SC with a defect point versus experimental
data for a complete SC and a SC with a point defect. a) Left
panel: Complex Band Structure calculated by EPWE with the
supercell approximation. Central panel: Real Band Structure. Green
line represents the localized mode. Right panel: Experimental
Insertion Loss of the complete SC (red line) in the Band Gap and
experimental Insertion Loss of the SC with a point defect (blue
line). The inset shows the supercell used in the calculations. Red
Square marks the value of the imaginary part of the wave vector
$Im(k)=-5.6$. b) Band structures for a complete SC. SC made of PVC
cylinders with $r=0.1$m, $\rho_{PVC}=1400$kg/$m^3$ and
$c_{PVC}=2380$m/s, embedded in air, $\rho_{air}=1.23$kg/$m^3$ and
$c_{air}=340$m/s.}
\end{figure}

Figure \ref{fig:Figure_1} shows the complex and real band
structures for the SC with a point defect. The complex band
structures and the value of the $k$ number for the modes inside
the BG can be obtained by EPWE and it becomes in a purely real
value for the localized mode. We can observe the localized mode at
$920$Hz (green continuous line). That value exactly coincides with
the value obtained by Plane Wave Expansion (PWE) with supercell
approximation. We have compared these results with experimental
data measuring the Insertion Loss (IL) behind the SC with and
without the point defect. In Figure \ref{fig:Figure_1} we can
observe that the experimental IL for the localized mode at
frequency $920$Hz (blue line) is lower than the case of the
complete SC (red line), i.e., it can be concluded that there is a
passing mode. This results because the localized mode is not
killed completely by the SC around the point defect (see also
\cite{Wu09}). In fact, although the localized mode has an
evanescent behaviour, as we will see later, in this case there is
not enough number of rows around the point defect to kill it.

\begin{figure}
\includegraphics[width=90mm,height=60mm,angle=0]{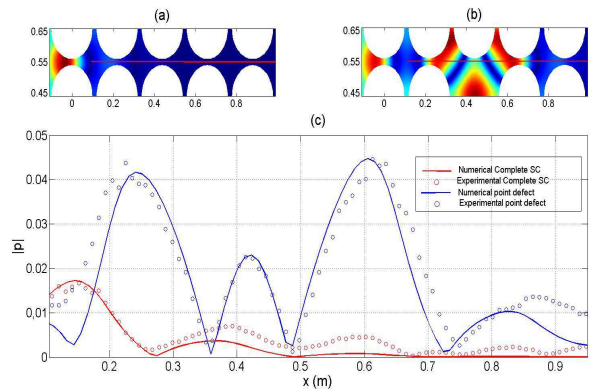}%
\caption{\label{fig:Figure_2}(Color online) Absolute values of the
acoustic field inside the SC with and without point defect.
Numerical maps calculated by FEM inside the complete SC (a) and
inside the SC with a point defect (b). (c) Numerical and
Experimental data for the the interior path marked in (a) and (b)
with a red line. Red continuous line (Red open circles) represents
the numerical (experimental) results for complete SC. Blue
continuous line (Blue open circles) represents the numerical
(experimental) results for SC with a point defect.}
\end{figure}

For frequencies in the BG, the borders of the point defect act as
perfect mirrors producing the localization in this
cavity\cite{Joannopoulus08}. The localized mode sees a complete SC
from inside the point defect in every directions. Thus, the
localized mode should present out of the cavity a decay analogous
to a wave with the same frequency impinging over the SC from
outside. Both cases should be represented by the same $Im(k)$,
i.e., by the same evanescent behaviour. There are several values
of the imaginary part of the wave vector at the localized
frequency in the Complex Band structures shown in Figure
\ref{fig:Figure_1} which shows the multiexponential behaviour. In
this paper we show for the localized frequency, 920Hz, that the
first harmonic obtained from the Figure \ref{fig:Figure_1} (red
square), $Im(k)=-5.6$m$^{-1}$, can represent thee decay of the
mode inside the SC. The contribution to the next harmonics to the
multiexponential decay can be neglected.

In order to study this behaviour of the localized mode we have
analyzed numerically the acoustic field inside the SC. In Figures
\ref{fig:Figure_2}a and \ref{fig:Figure_2}b we can observe the
maps obtained by Finite Element Method (FEM) for the complete SC
and for the SC with a point defect respectively. In Figure
\ref{fig:Figure_2}c we represent both numerical and experimental
absolute values of the pressure for complete SC and for SC with a
point defect corresponding to the cross sections marked with a red
line in Figures \ref{fig:Figure_2}a and \ref{fig:Figure_2}b.
Experimental results are also plotted in Figure
\ref{fig:Figure_2}c.

For a complete SC, we can observe in Figure \ref{fig:Figure_2}
both numerically (red continuous) and experimentally (red open
circles) exponential-like decay of the mode with the distance all
along the SC without point defect. From these experimental data,
we have chosen the points with maximum values in order to fit an
exponential $ae^{bx}$. The values of the parameters in the fit are
$a=0.02938\pm0.0103$ and $b=Im(k)=-5.60\pm1.45$m$^{-1}$.

In the blue line in Figure \ref{fig:Figure_2}c we can observe the
effect of the point defect in the acoustic field inside the SC. In
the region of the point defect there is an increasing value of the
acoustic pressure because of the localized mode. We can also
observe that the absolute value of the pressure for the localized
mode is bigger than in the case of the complete SC at the end of
the SC, proving the passing mode shown in Figure
\ref{fig:Figure_1}. To enhance the localization of the sound
inside the SC we would need a SC with a bigger number of rows
around the point defect as it was shown by some
authors\cite{Wu09}.

\begin{figure}
\includegraphics[width=80mm,height=70mm,angle=0]{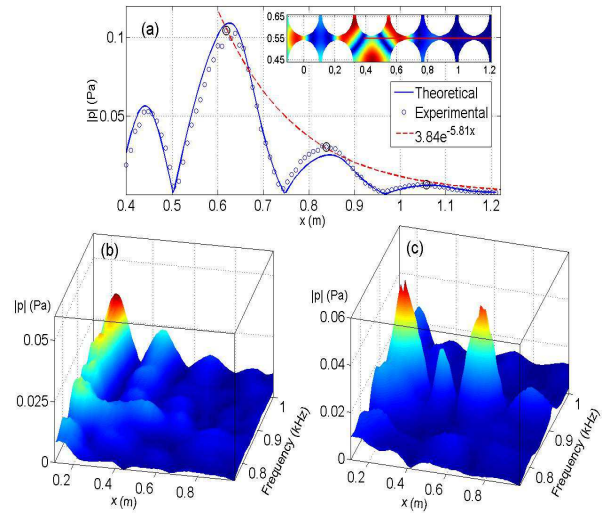}%
\caption{\label{fig:Figure_3}(Color online) (a) Absolute values of
pressure inside a 6$\times$5 SC with a point defect: Numerical
results (Blue line), experimental results (blue open circles). Red
dashed line represents the fitted exponential-like decay of the
localized mode using the black open circles. (b) and (c) represent
the 3D spectra for the complete SC and for the SC with a point
defect respectively.}
\end{figure}

The border of the cavity is located approximately at $x=0.6$m as
we can observe in Figure \ref{fig:Figure_2}b. From this point to
the end of the SC we can observe that the acoustic field is
drastically reduced, but with these evidences we cannot confirm
that the behaviour of the localized mode out of the cavity is
evanescent. To do that, we have analyzed the sound inside a bigger
SC with a point defect (see inset of Figure \ref{fig:Figure_3}a).
Figure \ref{fig:Figure_3}a presents both numerical (blue line) and
experimental (blue open circles) values of the acoustic field from
the end of the cavity to the end of a SC showing the evanescent
behaviour of the localized mode out of the cavity. Analogously as
the case of complete SC, here we have also chosen the points with
maximum values (see black open circles in Figure
\ref{fig:Figure_3}a) in order to fit an exponential decay
$ae^{bx}$. The values of the parameters in the fit are
$a=3.84\pm9.92$ and $b=Im(k)=-5.81\pm4.06$m$^{-1}$ and the curve
is also plotted in Figure \ref{fig:Figure_3}a with red dashed
line. From the experimental point of view we are constrained by
the size of the SC, and as a consequence we have only been able to
use a few points for the exponential fit. This results in a big
error in the parameters of the fit. Even so, the value obtained
for the $Im(k)$ is very closed to the one obtained both
analytically (EPWE) and experimentally for the complete SC. The
difference is less than 4\% in both cases.

We have analyzed one frequency ($920$Hz) inside the BG, but this
exponential-like decay should be observed for every the
frequencies inside BG independently if there is or not a point
defect in the SC. Figures \ref{fig:Figure_3}b and
\ref{fig:Figure_3}c represent the experimental 3D spectra for both
complete SC and SC with a point defect respectively. %The spectrum
%at a point $x_0$ is plotted in the YZ-plane, such that the
%frequency is in the y-axis and the absolute value of the pressure
%is in the z-axis. We construct the 3D spectra, plotting the
%spectra for all the points inside the SC between two rows
in the range of frequencies [750, 1000]Hz, in the BG. In Figure
\ref{fig:Figure_3}b we observe the experimental evidences of the
evanescent behaviour for all the modes inside the BG for a
complete SC. For the case of the SC with a point defect there is a
change in the propagation properties. We observe clearly in Figure
\ref{fig:Figure_3}c the evanescent behaviour of the localized
frequencies out of the region of the point defect and how the
localization is produced, showing the effect of the
cavity\cite{Wu09}. We can also observe in Figure
\ref{fig:Figure_3}c the evanescent behaviour for all modes out of
the frequencies of the localization range.

The propagation of waves inside periodic structures consists on
both propagating and evanescent modes. Using EPWE we have observed
the evanescent nature of the modes inside the BG with negative
complex Bloch vectors. From the experimental point of view we have
observed that the exponential-like decay is dominated by the first
harmonic of the Fourier expansion of the Bloch wave, obtaining
this exponential-like decay for the modes in the BG
($Im(k)=-5.6\pm1.45$m$^{-1}$ for $920$Hz) and for localized modes
in a SC with a point defect ($Im(k)=-5.81\pm4.06$m$^{-1}$). From
analytical and experimental data we can conclude that localized
modes present evanescent behaviour out of the cavity with the same
exponential-like decay than a wave with the same frequency
impinging over a complete SC. Due to the breaking periodicity, the
physical situations are very different between complete SC and SC
with point defects, even so we can conclude that the space
observed by the localized wave from inside of the cavity is
topologically equivalent to the observed by the same wave from
outside of a complete SC. This work is fundamental for the correct
understanding of the design of narrow filters and wave guides
based on Phononic Crystals with point defects.

\begin{acknowledgments}
This work was supported by MEC (Spanish Government) and FEDER
funds, under grands MAT2009-09438 and MTM2009-14483-C02-02.
Authors want to thanks J. M. Herrero, S. Garc\'ia-Nieto, and X.
Blasco for their work in the control and acquisition system of
3DReAMS.
\end{acknowledgments}

%\bibliography{Evanescent_PRL.bib}

\begin{thebibliography}{}%
\makeatletter
\providecommand \@ifxundefined [1]{%
 \ifx #1\undefined \expandafter \@firstoftwo
 \else \expandafter \@secondoftwo
\fi
}%
\providecommand \@ifnum [1]{%
 \ifnum #1\expandafter \@firstoftwo
 \else \expandafter \@secondoftwo
\fi
}%
\providecommand \enquote [1]{``#1''}%
\providecommand \bibnamefont  [1]{#1}%
\providecommand \bibfnamefont [1]{#1}%
\providecommand \citenamefont [1]{#1}%
\providecommand\href[0]{\@sanitize\@href}%
\providecommand\@href[1]{\endgroup\@@startlink{#1}\endgroup\@@href}%
\providecommand\@@href[1]{#1\@@endlink}%
\providecommand \@sanitize [0]{\begingroup\catcode`\&12\catcode`\#12\relax}%
\@ifxundefined \pdfoutput {\@firstoftwo}{%
 \@ifnum{\z@=\pdfoutput}{\@firstoftwo}{\@secondoftwo}%
}{%
 \providecommand\@@startlink[1]{\leavevmode}%
 \providecommand\@@endlink[0]{}%
}{%
 \providecommand\@@startlink[1]{%
  \leavevmode
  \pdfstartlink
   attr{/Border[0 0 1 ]/H/I/C[0 1 1]}%
   user{/Subtype/Link/A<</Type/Action/S/URI/URI(#1)>>}%
  \relax
 }%
 \providecommand\@@endlink[0]{\pdfendlink}%
}%
\providecommand \url  [0]{\begingroup\@sanitize \@url }%
\providecommand \@url [1]{\endgroup\@href {#1}{\urlprefix}}%
\providecommand \urlprefix [0]{URL }%
\providecommand \Eprint[0]{\href }%
\@ifxundefined \urlstyle {%
  \providecommand \doi [1]{doi:\discretionary{}{}{}#1}%
}{%
  \providecommand \doi [0]{doi:\discretionary{}{}{}\begingroup
  \urlstyle{rm}\Url }%
}%
\providecommand \doibase [0]{http://dx.doi.org/}%
\providecommand \Doi[1]{\href{\doibase#1}}%
\providecommand \bibAnnote [3]{%
  \BibitemShut{#1}%
  \begin{quotation}\noindent
    \textsc{Key:}\ #2\\\textsc{Annotation:}\ #3%
  \end{quotation}%
}%
\providecommand \bibAnnoteFile [2]{%
  \IfFileExists{#2}{\bibAnnote {#1} {#2} {\input{#2}}}{}%
}%
\providecommand \typeout [0]{\immediate \write \m@ne }%
\providecommand \selectlanguage [0]{\@gobble}%
\providecommand \bibinfo [0]{\@secondoftwo}%
\providecommand \bibfield [0]{\@secondoftwo}%
\providecommand \translation [1]{[#1]}%
\providecommand \BibitemOpen[0]{}%
\providecommand \bibitemStop [0]{}%
\providecommand \bibitemNoStop [0]{.\EOS\space}%
\providecommand \EOS [0]{\spacefactor3000\relax}%
\providecommand \BibitemShut [1]{\csname bibitem#1\endcsname}%
%</preamble>
\bibitem{Sigalas93}%
  \BibitemOpen
  \bibfield{author}{%
  \bibinfo {author} {\bibfnamefont{M.}~\bibnamefont{Sigalas}}\ and\ \bibinfo
  {author} {\bibfnamefont{E.}~\bibnamefont{Economou}},\ }%
  \bibfield{journal}{%
  \bibinfo {journal} {Solid State Commun.}\ }%
  \textbf{\bibinfo {volume} {86}},\ \bibinfo {pages} {141} (\bibinfo {year}
  {1993})%
  \bibAnnoteFile{NoStop}{Sigalas93}%
\bibitem{Kushwaha94PRB}%
  \BibitemOpen
  \bibfield{author}{%
  \bibinfo {author} {\bibfnamefont{M.~S.}~\bibnamefont{Kushwaha}}, \bibinfo
  {author} {\bibfnamefont{P.}~\bibnamefont{Halevi}}, \bibinfo {author}
  {\bibfnamefont{G.}~\bibnamefont{Mart\'inez}}, \bibinfo {author}
  {\bibfnamefont{L.}~\bibnamefont{Dobrzynski}},\ and\ \bibinfo {author}
  {\bibfnamefont{B.}~\bibnamefont{Djafari-Rouhani}},\ }%
  \bibfield{journal}{%
  \bibinfo {journal} {Phys. Rev. B}\ }%
  \textbf{\bibinfo {volume} {49}},\ \bibinfo {pages} {2313} (\bibinfo {year}
  {1994})%
  \bibAnnoteFile{NoStop}{Kushwaha94PRB}%
\bibitem{Yablonovitch}%
  \BibitemOpen
  \bibfield{author}{%
  \bibinfo {author} {\bibfnamefont{E.}~\bibnamefont{Yablonovitch}},\ }%
  \bibfield{journal}{%
  \bibinfo {journal} {Phys. Rev. Lett.}\ }%
  \textbf{\bibinfo {volume} {58}},\ \bibinfo {pages} {2059} (\bibinfo {year}
  {1987})%
  \bibAnnoteFile{NoStop}{Yablonovitch}%
\bibitem{Martinez95}%
  \BibitemOpen
  \bibfield{author}{%
  \bibinfo {author} {\bibfnamefont{R.}~\bibnamefont{Mart\'inez-Sala}}, \bibinfo
  {author} {\bibfnamefont{J.}~\bibnamefont{Sancho}}, \bibinfo {author}
  {\bibfnamefont{J.~V.}\ \bibnamefont{S\'anchez}}, \bibinfo {author}
  {\bibfnamefont{V.}~\bibnamefont{G\'omez}}, \bibinfo {author}
  {\bibfnamefont{J.}~\bibnamefont{Llinares}},\ and\ \bibinfo {author}
  {\bibfnamefont{F.}~\bibnamefont{Meseguer}},\ }%
  \bibfield{journal}{%
  \bibinfo {journal} {nature}\ }%
  \textbf{\bibinfo {volume} {378}},\ \bibinfo {pages} {241} (\bibinfo {year}
  {1995})%
  \bibAnnoteFile{NoStop}{Martinez95}%
\bibitem{Sanchez98}%
  \BibitemOpen
  \bibfield{author}{%
  \bibinfo {author} {\bibfnamefont{J.~V.}\ \bibnamefont{S\'anchez-P\'erez}},
  \bibinfo {author} {\bibfnamefont{D.}~\bibnamefont{Caballero}}, \bibinfo
  {author} {\bibfnamefont{R.}~\bibnamefont{Mart\'inez-Sala}}, \bibinfo {author}
  {\bibfnamefont{C.}~\bibnamefont{Rubio}}, \bibinfo {author}
  {\bibfnamefont{J.}~\bibnamefont{S\'anchez-Dehesa}}, \bibinfo {author}
  {\bibfnamefont{F.}~\bibnamefont{Meseguer}}, \bibinfo {author}
  {\bibfnamefont{J.}~\bibnamefont{Llinares}},\ and\ \bibinfo {author}
  {\bibfnamefont{F.}~\bibnamefont{G\'alvez}},\ }%
  \bibfield{journal}{%
  \bibinfo {journal} {Phys. Rev. Lett.}\ }%
  \textbf{\bibinfo {volume} {80}},\ \bibinfo {pages} {5325} (\bibinfo {year}
  {1998})%
  \bibAnnoteFile{NoStop}{Sanchez98}%
\bibitem{Joannopoulus08}%
  \BibitemOpen
  \bibfield{author}{%
  \bibinfo {author} {\bibfnamefont{J.~D.}\ \bibnamefont{Joannopoulus}},
  \bibinfo {author} {\bibfnamefont{S.~G.}\ \bibnamefont{Johnson}}, \bibinfo
  {author} {\bibfnamefont{J.~N.}\ \bibnamefont{Winn}},\ and\ \bibinfo {author}
  {\bibfnamefont{R.~D.}\ \bibnamefont{Meade}},\ }%
  \emph{\bibinfo {title} {Photonic Crystals. Molding the Flow of Light}}\
  (\bibinfo {publisher} {Princeton University press}, \bibinfo {city} {Princeton},\ \bibinfo {year} {2008})%
  \bibAnnoteFile{NoStop}{Joannopoulus08}%
\bibitem{Torrent06}%
  \BibitemOpen
  \bibfield{author}{%
  \bibinfo {author} {\bibfnamefont{D.}\ \bibnamefont{Torrent}},
  \bibinfo {author} {\bibfnamefont{A.}\ \bibnamefont{Hakansson}}, \bibinfo
  {author} {\bibfnamefont{F.}\ \bibnamefont{Cervera}},\ and\ \bibinfo {author}
  {\bibfnamefont{J.}\ \bibnamefont{S\'anchez-Dehesa}},\ }%
  \bibfield{journal}{%
  \bibinfo {journal} {Phys. Rev. Lett.}\ }%
  \textbf{\bibinfo {volume} {96}},\ \bibinfo {pages} {204302} (\bibinfo {year}
  {2006})%
  \bibAnnoteFile{NoStop}{Torrent06}%
\bibitem{Torrent07}%
  \BibitemOpen
  \bibfield{author}{%
  \bibinfo {author} {\bibfnamefont{D.}\ \bibnamefont{Torrent}},
  \bibinfo {author} {\bibfnamefont{J.}\ \bibnamefont{S\'anchez-Dehesa}},\ }%
  \bibfield{journal}{%
  \bibinfo {journal} {New. Jour. Phys.}\ }%
  \textbf{\bibinfo {volume} {9}},\ \bibinfo {pages} {323} (\bibinfo {year}
  {2007})%
  \bibAnnoteFile{NoStop}{Torrent07}%
\bibitem{Wu01}%
  \BibitemOpen
  \bibfield{author}{%
  \bibinfo {author} {\bibfnamefont{F.}~\bibnamefont{Wu}}, \bibinfo {author}
  {\bibfnamefont{Z.}~\bibnamefont{Hou}}, \bibinfo {author}
  {\bibfnamefont{Z.}~\bibnamefont{Liu}},\ and\ \bibinfo {author}
  {\bibfnamefont{Y.}~\bibnamefont{Liu}},\ }%
  \bibfield{journal}{%
  \bibinfo {journal} {Phys. Lett. A}\ }%
  \textbf{\bibinfo {volume} {292}},\ \bibinfo {pages} {198} (\bibinfo {year}
  {2001})%
  \bibAnnoteFile{NoStop}{Wu01}%
\bibitem{Wu09}%
  \BibitemOpen
  \bibfield{author}{%
  \bibinfo {author} {\bibfnamefont{L.}~\bibnamefont{Wu}}, \bibinfo {author}
  {\bibfnamefont{L.}~\bibnamefont{Chen}},\ and\ \bibinfo {author}
  {\bibfnamefont{C.}~\bibnamefont{Liu}},\ }%
  \bibfield{journal}{%
  \bibinfo {journal} {Physica B}\ }%
  \textbf{\bibinfo {volume} {404}},\ \bibinfo {pages} {1766} (\bibinfo {year}
  {2009})%
  \bibAnnoteFile{NoStop}{Wu09}%
\bibitem{Vasseur08}%
  \BibitemOpen
  \bibfield{author}{%
  \bibinfo {author} {\bibfnamefont{J.~O.}\ \bibnamefont{Vasseur}}, \bibinfo
  {author} {\bibfnamefont{P.~A.}\ \bibnamefont{Deymier}}, \bibinfo {author}
  {\bibfnamefont{B.}~\bibnamefont{Djafari-Rouhani}}, \bibinfo {author}
  {\bibfnamefont{Y.}~\bibnamefont{Pennec}},\ and\ \bibinfo {author}
  {\bibfnamefont{A.C.}\ \bibnamefont{Hladky-Hennion}},\ }%
  \bibfield{journal}{%
  \bibinfo {journal} {Phys. Rev.B}\ }%
  \textbf{\bibinfo {volume} {77}},\ \bibinfo {pages} {085415} (\bibinfo {year}
  {2008})%
  \bibAnnoteFile{NoStop}{Vasseur08}%
\bibitem{Engelen09}%
  \BibitemOpen
  \bibfield{author}{%
  \bibinfo {author} {\bibfnamefont{R.}~\bibfnamefont{J.}~\bibfnamefont{P.}~\bibnamefont{Engelen}}, \bibinfo
  {author} {\bibfnamefont{D.}~\bibnamefont{Mori}}, \bibinfo {author}
  {\bibfnamefont{T.}~\bibnamefont{Baba}},\ and\ \bibinfo {author}
  {\bibfnamefont{L.}~\bibnamefont{Kuipers}},\ }%
  \bibfield{journal}{%
  \bibinfo {journal} {Phys. Rev. Lett.}\ }%
  \textbf{\bibinfo {volume} {102}},\ \bibinfo {pages} {023902} (\bibinfo {year}
  {2009})%
  \bibAnnoteFile{NoStop}{Engelen09}%
\bibitem{Laude09}%
  \BibitemOpen
  \bibfield{author}{%
  \bibinfo {author} {\bibfnamefont{V.}\ \bibnamefont{Laude}},
  \bibinfo {author} {\bibfnamefont{Y.}\ \bibnamefont{Achaoui}},
  \bibinfo {author} {\bibfnamefont{S.}\ \bibnamefont{Benchabane}},\ and\ \bibinfo {author}
  {\bibfnamefont{A.}\ \bibnamefont{Khelif}},\ }%
  \bibfield{journal}{%
  \bibinfo {journal} {Phys. Rev. B}\ }%
  \textbf{\bibinfo {volume} {80}},\ \bibinfo {pages} {092301} (\bibinfo {year}
  {2009})%
  \bibAnnoteFile{NoStop}{Laude09}%
\bibitem{Romero10}%
  \BibitemOpen
  \bibfield{author}{%
  \bibinfo {author} {\bibfnamefont{V.}~\bibnamefont{Romero-Garc\'ia}}, \bibinfo {author}
  {\bibfnamefont{J.V.}~\bibnamefont{S\'anchez-P\'erez}}, \ and\ \bibinfo {author}
  {\bibfnamefont{L.M.}~\bibnamefont{Garcia-Raffi}},\ }%
  \bibfield{journal}{%
  \bibinfo {journal} {Preprint, arXiv:1001.3758v1}\ }%
(\bibinfo {year}
  {2010})%
  \bibAnnoteFile{NoStop}{Romero10}%
\bibitem{Hsue05}%
  \BibitemOpen
  \bibfield{author}{%
  \bibinfo {author} {\bibfnamefont{Young-Chung}~\bibnamefont{Hsue}}, \bibinfo {author}
  {\bibfnamefont{Arthur J.}~\bibnamefont{Freeman}},\ and\ \bibinfo {author}
  {\bibfnamefont{Ben-Yuan}~\bibnamefont{Gu}},\ }%
  \bibfield{journal}{%
  \bibinfo {journal} {Phys. Rev B}\ }%
  \textbf{\bibinfo {volume} {72}},\ \bibinfo {pages} {195118} (\bibinfo {year}
  {2005})%
  \bibAnnoteFile{NoStop}{Hsue05}%
\bibitem{Khelif03}%
  \BibitemOpen
  \bibfield{author}{%
  \bibinfo {author} {\bibfnamefont{A.}~\bibnamefont{Khelif}}, \bibinfo {author}
  {\bibfnamefont{A.}~\bibnamefont{Choujaa}}, \bibinfo {author}
  {\bibfnamefont{B.}~\bibnamefont{Djafari-Rouhani}}, \bibinfo {author}
  {\bibfnamefont{M.}~\bibnamefont{Wilm}}, \bibinfo {author}
  {\bibfnamefont{S.}~\bibnamefont{Ballandras}},\ and\ \bibinfo {author}
  {\bibfnamefont{V.}~\bibnamefont{Laude}},\ }%
  \bibfield{journal}{%
  \bibinfo {journal} {Phys. Rev. B}\ }%
  \textbf{\bibinfo {volume} {68}},\ \bibinfo {pages} {214301} (\bibinfo {year}
  {2003})%
  \bibAnnoteFile{NoStop}{Khelif03}%
%\bibitem{Zhao09}%
%  \BibitemOpen
%  \bibfield{author}{%
%  \bibinfo {author} {\bibfnamefont{Y.}~\bibnamefont{Zhao}}\ and\ \bibinfo
%  {author} {\bibnamefont{L.B.Yuan}},\ }%
%  \bibfield{journal}{%
%  \bibinfo {journal} {J. Phys. D: Appl. Phys.}\ }%
%  \textbf{\bibinfo {volume} {42}},\ \bibinfo {pages} {015403} (\bibinfo {year}
%  {2009})%
%  \bibAnnoteFile{NoStop}{Zhao09}%
%\bibitem{Sigalas98}%
%  \BibitemOpen
%  \bibfield{author}{%
%  \bibinfo {author} {\bibfnamefont{M.}~\bibnamefont{Sigalas}},\ }%
%  \bibfield{journal}{%
%  \bibinfo {journal} {J. Appl. Phys.}\ }%
%  \textbf{\bibinfo {volume} {84}},\ \bibinfo {pages} {3026} (\bibinfo {year}
%  {1998})%
%  \bibAnnoteFile{NoStop}{Sigalas98}%
\end{thebibliography}
%Merlin.mbs v4.21 2009-07-09.
%

\end{document}